\pdfoutput=1

\RequirePackage{snapshot}
\documentclass[12pt]{article}

\usepackage[T1]{fontenc}
\usepackage[utf8]{inputenc}

\usepackage
[a4paper,tmargin=3truecm,bmargin=3truecm,rmargin=2.5truecm,lmargin=2.5truecm,twoside,verbose=true]{geometry}
\usepackage{amsmath,amssymb,latexsym}

\usepackage{stmaryrd,wasysym,upgreek,mathrsfs,dsfont}
\usepackage[english]{babel}
\usepackage{graphicx,color}
\usepackage{slashed,subfig}
\usepackage[vflt]{floatflt}

\allowdisplaybreaks[1]

\usepackage[thmmarks,amsmath]{ntheorem}
\usepackage{fabmesthm-en}
\usepackage{fabmacromath}

\newcommand{\Gsub}{\sub{\Gamma}}

\numberwithin{equation}{section}

\usepackage{fabfeynmp} %inclut les figures filenameMpost-?.pdf a la place de filename.?
\title{Hopf algebra of non-commutative field theory}
\author{\textrm{Adrian Tanasa${}^{(1),(2)}$ and Fabien Vignes-Tourneret${}^{(3)}$}}
\date{}

\begin{document}
\maketitle

\vspace*{-1cm}
\begin{center}
\textrm{$^1$Laboratoire de Physique Théorique, Bât.\ 210\\
    Université Paris XI,  F-91405 Orsay Cedex, France\\
    $^2$Dep. Fizica Teoretica, Institul de Fizica si Inginerie Nucleara
    H. Hulubei,\\
    P. O. Box MG-6, 077125 Bucuresti-Magurele, Romania\\
    e-mail: \textsf{adrian.tanasa@ens-lyon.org}}\\
  \bigskip
  \textrm{$^3$IHÉS\\Le Bois-Marie, 35 route de Chartres\\F-91440 Bures-sur-Yvette, France\\
    e-mail: \textsf{vignes@ihes.fr}}
 \end{center}%

\begin{abstract}
We contruct here the Hopf algebra structure underlying the process of
renormalization of non-commutative quantum field theory.
\end{abstract}

\begin{fmffile}{ckMpost}

\fmfset{wiggly_len}{5pt} % reducing length of wiggles in boson lines
\fmfset{wiggly_slope}{70} % increasing slope of wiggles in boson lines
%\fmfset{curly_len}{4pt} % reducing length of wiggles in boson lines
\fmfset{curly_len}{1.4mm}

\fmfset{dot_len}{1mm}

% \begin{figure}[t]
% \begin{center}
% $S\big( $
%   \parbox{40pt}{
% \begin{fmfgraph*}(40,50)
%   \fmfleft{l}
%   \fmfright{r}
%   \fmf{photon}{l,v,r}
%   \fmfv{decor.shape=circle, decor.filled=0, decor.size=5thick}{v}
%   \end{fmfgraph*}
% }$~\big)=2$
% \hspace{3cm}
% $S\big( $
% \parbox{40pt}{
% \begin{fmfgraph*}(40,50)
%       \fmfleft{l}
%       \fmfright{r}
%       \fmf{plain}{l,v1,v2,r}
%       \fmf{photon,left,tension=0}{v1,v2}
% \end{fmfgraph*}
% }$~\big)=1$
% \end{center}
% \caption{Automorphisms of Feynman graphs respect the type of vertex/edge in $R$. }
% \end{figure}

\section{Introduction and motivation}

Hopf algebras (see for example  \cite{Kassel:1995qy} or
\cite{Dascualescu:2000uq}) are today one of the most studied structures in
mathematics. In relation with quantum field theories (QFT), Hopf algebras were
proven to be a natural framework for the description of the forest structure
of renormalization - the Connes-Kreimer algebras \cite{Connes:2000uq,Connes:2001kx}. Ever since 
 there has been an important amount of work with respect to this new class of
 Hopf algebras (for a general review see for example \cite{Kreimer:2005lr}).

\begin{floatingfigure}[p]{2cm}
  \centering
  \includegraphics[scale=1.3]{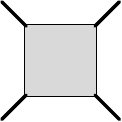}
  \caption[A Moyal vertex]{A Moyal vertex}
  \label{fig:vertex}
\end{floatingfigure}
\noindent
However, this construction was realized so far only at the level of commutative
QFT. When uplifting to non-commutative quantum field theory (NCQFT), the
interaction is no longer local. Thus,  the vertices of the associated Feynman
diagrams can now be represented as in Fig \ref{fig:vertex}.

Recently, NCQFT models were also proven to be renormalizable at any order in perturbation
theories, despite the ultraviolet-infrared mixing problem% \cite{MiRaSe}
. The non-commutative
analogous of the Bogoliubov-Parasiuk-Hepp-Zimmerman (BPHZ) theorem was proven for the Grosse-Wulkenhaar $\Phi^4$ scalar model in \cite{GrWu03-1,GrWu04-3}. In \cite{xphi4-05} a general proof in $x$-space, using
multiscale analysis was given. The parametric representation
was implemented for this model in \cite{gurauhypersyman}. 
Furthermore, the Mellin representation of the
non-commutative Feynman amplitudes was achieved in \cite{Guruau:2007fj}. Finally, 
 the dimensional regularization and
renormalization were constructed in \cite{Guruau:2007kx}. 

With respect to the form of the associated propagator, a second class of NCQFT
models exists. This second class contains the non-commutative Gross-Neveu and
the Langmann-Szabo-Zarembo \cite{Langmann2003if} models. The associate BPHZ
theorem was proven in \cite{RenNCGN05} for the non-commutative Gross-Neveu model. Moreover, the parametric representation \cite{Tanasa:2007fk} and
the Mellin representation \cite{Guruau:2007fj} were also implemented for this class of
models too. For a recent 
review on different issues of renormalizability of NCQFT the interested
reader may report himself to \cite{Rivasseau:2007yq}.

Note that even though recent progress has been made in \cite{De-Goursac:2006lr,Grosse:2007vn,Blaschke:2007rt}, physicists do not yet have a renormalizable non-commutative gauge theory.

\medskip

In this article we construct the Hopf algebra structure  associated to the renormalization of these NCQFT models.
The paper is organized as follows. 
In the next section we give some insights on the renormalization of NCQFT with
respect to renormalization of commutative QFT. The third section is
devoted to the Hopf algebra structure of Feynman diagrams. In the last section
we state and prove our main result.
%, namely the implementation of a Hopf algebra
%structure  related to renormalization of NCQFT. 

\section{Renormalization of non-commutative quantum field theory}

% %% À partir d'ici, pdfdraftcopy n'imprime plus rien sur les pages
% \ClearDraftOverlay

In this section we briefly recall some features of both commutative and
non-commutative Euclidean renormalization. We will mainly focus on the Grosse-Wulkenhaar model \cite{GrWu04-3,xphi4-05} or non-commutative $\Phi^{4}_{4}$ theory. It consists in a scalar quantum field theory on the four-dimensional Moyal space. Its action is given by
\begin{equation}\label{action}
S[\phi] = \int d^4x \Big( -\frac{1}{2} \phi(-\Delta)\phi + \frac{\Omega^2}{2}\xt^{2}\phi^{2}+ \frac{1}{2} m^2\,\phi^{2}
+ \frac{\lambda}{4} \phi \star \phi \star \phi \star
\phi\Big)(x)
\end{equation}
with $\xt_\mu=2(\Theta^{-1}x)_{\mu}$ and $\Theta$ a four by four
skew-symmetric matrix which encodes the non-commutative character of space
time: $[x^\mu, x^\nu]_{\star}=\imath\Theta^{\mu\nu}$. It has been shown renormalizable to all orders of perturbation.

Furthermore, as already stated in the previous section, the same renormalization results also hold for the \encv{} Gross-Neveu model \cite{RenNCGN05} and the \emph{generalized} LSZ model \cite{xphi4-05}:
\begin{align}
    S_{\text{GN}}=&\int d^2x\big[{\bar{\psi}}(-i{\slashed{\partial}}+\Omega\xts+m+\mu\gamma_5)\psi-
    \sum_{A=1}^3\frac{g_A}{4}({\cal{J}}^{A}\star{\cal{J}}^{A})(x)\big], \label{eq:actionGN}\\
    {\cal{J}}^{A}=&{\bar{\psi}}\star\Gamma^{A}\psi,\ \Gamma_{1}=\bbbone,\,\Gamma_{2}=\gamma^{\mu},\,\Gamma_{3}=\gamma_{5},\label{eq:courants},\\
    S_{\text{gLSZ}}=&\int d^{4}x\big[\bar{\phi}\big((-\imath\partial_{\mu}+\Omega_{1}\xt_{\mu})^{2}+\Omega_{2}\xt^{2}+m^{2}\big)\phi+\frac{\lambda}{2}\bar{\phi}\star\phi\star\bar{\phi}\star\phi\big](x).\label{eq:actiongLSZ}
  \end{align}

\subsection{Topology and power counting}
\label{sec:topol-feynm-graphs}

Let a graph G with $V$ vertices and $I$ internal lines. Interactions of
quantum field theories on the Moyal space are only invariant under \textbf{cyclic permutation} of the incoming/outcoming fields. This restriced
invariance replaces the permutation invariance which was present in the case of local interactions.

A good way to keep track of such a reduced invariance is to draw Feynman graphs as ribbon graphs. Moreover there exists a basis for the Schwartz class functions where the Moyal product becomes an ordinary matrix product \cite{GrWu03-2,Gracia-Bondia1987kw}. This further justifies the ribbon representation.\\

\begin{figure}[htbp]
  \centering 
  \subfloat[$x$-space representation]{{\label{fig:x-rep}}\includegraphics[scale=.8]{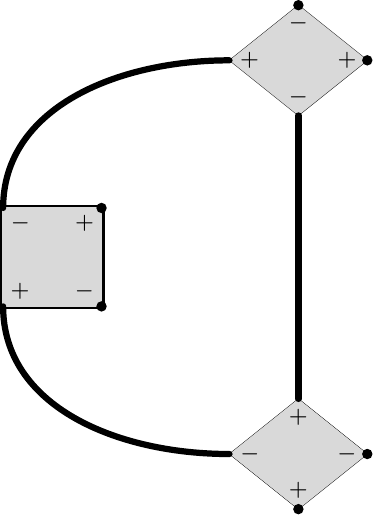}}\hspace{2cm}\qquad
  \subfloat[Ribbon representation]{\label{fig:ribbon}\includegraphics[scale=1]{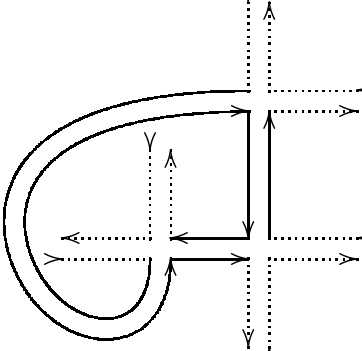}}
  \caption{A graph with two broken faces}
  \label{fig:broken-ex}
\end{figure}

Let us consider the example of figure \ref{fig:broken-ex}.
Propagators in a ribbon graph are made of double lines. Let us call $F$ the number of faces (loops made of single lines) of a ribbon graph. The graph of figure \ref{fig:ribbon} has $V=3, I=3, F=2$. Each ribbon graph can be drawn on a manifold of genus $g$. The genus is computed from the Euler characteristic $\chi=F-I+V=2-2g$. If $g=0$ one has a {\it planar graph}, otherwise one has a {\it non-planar graph}. For example, the graph of figure \ref{fig:ribbon} may be drawn on a manifold of genus $0$. Note that some of the $F$ faces of a graph may be ``broken'' by external legs. In our example, both faces are broken. We denote the number of broken faces by $B$.\\

Furthermore let $N$ the number of external legs of the graph. For the commutative $\phi^4$ model one has the following superficial degree of convergence $\omega= N-4$. Thus one has to deal only with the renormalization of the two- and four-point functions. In the case of the Grosse-Wulkenhaar model, the situation is different. In \cite{GrWu03-1,GrWu04-3,gurauhypersyman} it was proven that
\begin{equation}
\omega= (N-4)+8g+4(B-1).
\end{equation}
Note that, as proven in \cite{Tanasa:2007fk}  one has the same power counting for the LSZ like model (\ref{eq:actiongLSZ}). The one of the Gross-Neveu model (\ref{eq:actionGN}) is more involved but leads to the same conclusion: one has to deal only with the renormalization of the $B=1$, planar two- and four-point graphs hereafter qualified as \emph{planar regular}.

\subsection{Locality vs Moyality}
\label{sec:moyality}

A crucial aspect of the uplifting from commutative to non-commutative
renormalization is that the principle of locality of renormalized interactions of commutative QFT is
replaced with a new principle: renormalized interactions have a non-local
Moyal vertex form. This is nothing but the analog of the locality phenomenon which occurs in
commutative renormalization. One can thus speak, in the case of
non-commutative renormalization, of a new type of renormalization group, where
the locality is just replaced by ``Moyality''. The divergent parts of the planar regular two- and four-point graphs with one broken face (the only divergent graphs) are proportional to the ($1$PI) tree level terms of the perturbative expansion. Such a new definition of ``locality'' was suggested in \cite{Kreimer:2005lr}, see equation $(62)$.

Let us also argue here that, despite this uplifting, the combinatorial backbone
of renormalization theory is almost the same when dealing with commutative or
non-commutative QFT. Thus the combinatorics of non-commutative renormalization will be shown to be encoded by a Hopf algebra.

\subsection{Renormalization as a factorization issue}

The basic operation for renormalization is the disentanglement of a graph
$\Gamma$ into pieces $\gamma$ and cograph $\Gamma/\gamma$. It is exactly this
operation that was present at the level of commutative renormalization and
that gave rise to a Hopf algebra structure.

We now argue that this factorization process is also present at the level of
non-commutative renormalization.
Indeed, consider the dimensional renormalization scheme for the
Grosse-Wulkenhaar model. The parametric representation constructed in
\cite{gurauhypersyman} writes the Feynman amplitude $\phi (\Gamma)$ as
\begin{equation}
\label{HUGV}
\phi (\Gamma) = K  \int_{0}^{1} \prod_{\ell=1}^L  [ d t_\ell
(1-t_\ell^2)^{\frac D2 -1} ]
HU_{G, \bar{V}} ( t )^{-\frac D2}   
e^{-  \frac {HV_{G}}{HU_{G}}},
\end{equation}
\noindent
where $K$ is some constant, 
\begin{equation}
\label{t}
t_\ell = {\rm tanh} \frac{\alpha_\ell}{2}, \ \ell=1,\ldots, L,
\end{equation}
\noindent
where  $\alpha_\ell$
are the parameters associated to any of
the propagators of the graph. 
In \cite{gurauhypersyman} it was furthermore proved that  $HU$ and $HV$
are polynomials in the set of variables $t_\ell$.

Considering now a primitive divergent subgraph $\gamma$ of $\Gamma$ and
rescaling the parameters $t$ of its internal edges, it was proven in
\cite{Guruau:2007kx} that
\begin{equation}
\label{rezultat}
HU^{l}_\Gamma=HU^l_\gamma\, HU_{\Gamma/\gamma}
\end{equation}
where by the index $l$ we understand the leading terms under the rescaling. A
similar factorization theorem was also proven for the exponential part in
\eqref{HUGV} of the Feynman amplitude $\phi (\Gamma)$.

Moreover, in \cite{xphi4-05} an analogous phenomena of factorization was shown for
the Grosse-Wulkenhaar model in position space namely the planar regular graphs contribute to the renormalization of the mass, wave-function, harmonic frequency $\Omega$ and coupling constant, see Equation \eqref{action}.

\newpage
\section{Hopf algebra structure of Feynman diagrams}

\subsection{Hopf algebra reminder}

In this subsection we recall the general definition of a Hopf algebra (for further details one can refer
for example to \cite{Kassel:1995qy,Dascualescu:2000uq}).
\begin{defn}[Algebra]\label{defn:algebra}
  A unital associative algebra $\cA$ over a field $\K$ is a $\K$-linear space endowed with two algebra homomorphisms: 
  \begin{itemize}
  \item a product $m :\cA\otimes\cA\to\cA$ satisfying the \emph{associativity} condition:
    \begin{align}
      \forall\Gamma\in\cA,\,m\circ(m\otimes\id)(\Gamma)=&m\circ(\id\otimes m)(\Gamma),\label{eq:asso}
    \end{align}
  \item a unit $u :\K\to\cA$ satisfying:
    \begin{align}
      \forall \Gamma\in\cA,\,m\circ(u\otimes\id)(\Gamma)=&\Gamma=m\circ(\id\otimes u)(\Gamma).
    \end{align}
  \end{itemize}
\end{defn}

\begin{defn}[Coalgebra]\label{defn:coalgebra}
  A coalgebra $\cC$ over a field $\K$ is a $\K$-linear space endowed with two algebra homomorphisms: 
  \begin{itemize}
  \item a coproduct $\Delta :\cC\to\cC\otimes\cC$ satisfying the \emph{coassociativity} condition:
    \begin{align}
      \forall\Gamma\in\cC,\,(\Delta\otimes\id)\circ\Delta(\Gamma)=&(\id\otimes \Delta)\circ\Delta(\Gamma),\label{eq:coasso}
    \end{align}
  \item a counit $\veps :\cC\to\K$ satisfying:
    \begin{align}
      \forall \Gamma\in\cC,\,(\veps\otimes\id)\circ\Delta(\Gamma)=&\Gamma=(\id\otimes\veps)\circ\Delta(\Gamma).    \end{align}
  \end{itemize}
\end{defn}

\begin{defn}[Bialgebra]
  A bialgebra $\cB$ over a field $\K$ is a $\K$-linear space endowed with both an algebra and a coalgebra structure (see Definitions \ref{defn:algebra} and \ref{defn:coalgebra}) such that the coproduct and the counit are unital algebra homomorphisms (or equivalently the product and unit are coalgebra homomorphisms):
  \begin{subequations}
    \label{eq:compatibility}
    \begin{align}
      \Delta\circ m_{\cB}=&m_{\cB\otimes\cB}\circ(\Delta\otimes\Delta),\ \Delta(\bbbone)=\bbbone\otimes\bbbone,\\
      \veps\circ m_{\cB}=&m_{\K}\circ(\veps\otimes\veps),\ \veps(\bbbone)=1.
      \end{align}
    \end{subequations}
\end{defn}

\begin{defn}[Graded Bialgebra]
  A graded bialgebra is a bialgebra graded as a linear space: 
  \begin{align}
      \cB=\bigoplus_{n=0}^\infty\cB^{(n)}   
  \end{align}
  such that the grading is compatible with the algebra and coalgebra structures:
  \begin{align}
    \cB^{(n)}\cB^{(m)} \subseteq \cB^{(n+m)}\text{ and }\Delta\cB^{(n)}\subseteq\bigoplus_{k=0}^n\cB^{(k)}\otimes\cB^{(n-k)} .
  \end{align}
\end{defn}

\begin{defn}[Connectedness]
  A connected bialgebra is a graded bialgebra $\cB$ for which $\cB^{(0)}=u(\K)$.
\end{defn}
One can then define a Hopf algebra:
\begin{defn}[Hopf algebra]
  A Hopf algebra $\cH$ over a field $\K$ is a bialgebra over $\K$ equipped with an antipode map $S:\cH\to\cH$ obeying:
  \begin{align}
    m\circ(S\otimes\id)\circ\Delta=&u\circ\veps=m\circ(\id\otimes S)\circ\Delta.
  \end{align}
\end{defn}
Finally we remind a useful lemma:
\begin{lemma}[\cite{Manchon:2004fk}]\label{lem:Sfree}
  Any connected graded bialgebra is a Hopf algebra whose antipode is given by $S(\bbbone)=\bbbone$ and recursively by any of the two following formulas for $\Gamma\neq\bbbone$:
  \begin{subequations}
    \begin{align}
      S(\Gamma)=&-\Gamma-\sum_{(\Gamma)}S(\Gamma')\Gamma'',\label{eq:Srecurs}\\
      S(\Gamma)=&-\Gamma-\sum_{(\Gamma)}\Gamma'S(\Gamma'')
    \end{align}
  \end{subequations}
where we used Sweedler's notation.
\end{lemma}

\subsection{Locality and the residue map}
\label{sec:residue-map}

In quantum field theory, Feynman graphs are built from a certain set of edges and vertices $R=R_{E}\cup R_{V}$. This set is given by the particle content of the model and by the type of interactions one wants to consider. For example, in the commutative $\phi^{4}_{4}$ theory (which will be our benchmark until section \ref{sec:hopf-algebra-nonc}), $R_{E}$ contains only the scalar bosonic line while $R_{V}$ contains the local four-point vertex and the two-point vertices corresponding to the mass and wave-function renormalization:

\begin{align*}
R_E &= \{~
\parbox{20pt}{
    \begin{fmfgraph}(20,10)
      \fmfleft{l}
      \fmfright{r}
      \fmf{plain}{l,r}
    \end{fmfgraph}}~\},\;
R_V =\{~ 
\parbox{20pt}{
    \begin{fmfgraph}(20,10)
      \fmfleft{l1,l2}
      \fmfright{r1,r2}
      % \fmftop{t}
%       \fmfbottom{b}
      \fmf{plain}{l1,v,l2}
      \fmf{plain}{r1,v,r2}
      % \fmf{plain}{l1,vl1}
%       \fmf{plain}{l2,vl2}
%       \fmf{plain}{r1,vr1}
%       \fmf{plain}{r2,vr2}
      \fmfv{decor.shape=circle, decor.filled=full, decor.size=1thick}{v}
%      \fmfv{decor.shape=square, decor.filled=30, decor.size=4thick}{v}
      %\fmfpoly{filled=30,pull=1.0}{vl2,vl1,vr1,vr2}
      % \fmf{plain}{v,t}
%       \fmf{plain}{v,b}
%       \fmf{plain}{v,r}
    \end{fmfgraph}}~,~
  \parbox{20pt}{
    \begin{fmfgraph*}(20,10)
      \fmfleft{l}
      \fmfright{r}
      \fmf{plain}{l,v}
      \fmf{plain}{r,v}
      %\fmfdot{v}
      %\fmfblob{1.5thick}{v}
      \fmfv{decor.shape=circle, decor.filled=full, decor.size=1.0thick}{v}
      \fmfv{label={\tiny{$0$}},label.angle=90,label.dist=1.5mm}{v}
    \end{fmfgraph*}}~,~
\parbox{20pt}{
    \begin{fmfgraph*}(20,10)
      \fmfleft{l}
      \fmfright{r}
      \fmf{plain}{l,v}
      \fmf{plain}{r,v}
      %\fmfdot{v}
      %\fmfblob{1.5thick}{v}
      \fmfv{decor.shape=circle, decor.filled=full, decor.size=1.0thick}{v}
%       \fmfv{decor.shape=cross, decor.size=2.5thick}{v}
      \fmfv{label={\tiny{$1$}},label.angle=90,label.dist=1.5mm}{v}
    \end{fmfgraph*}}
~\}.
\end{align*}
In the following we will still write $R_{V}$ for the free algebra generated by the elements of $R_{V}$. Let us now consider the algebra $\cH$ generated by a certain class of graphs (connected, $1$PI etc) made out of the set $R$.
\begin{defn}[Subgraph]\label{def:subgraph}
  Let $\Gamma\in\cH$, $\Gamma^{[1]}$ its set of internal lines and $\Gamma^{[0]}$ its vertices. A \textbf{subgraph} $\gamma$ of $\Gamma$, written $\gamma\subset\Gamma$, consists in a subset $\gamma^{[1]}$ of $\Gamma^{[1]}$ and the vertices of $\Gamma^{[0]}$ hooked to the lines in $\gamma^{[1]}$. Note that with such a definition, $\gamma$ is truncated.
\end{defn}
\begin{defn}[Shrinkable subgraph]\label{def:shrinksubgraph}
  Let $\Gamma\in\cH$. A subgraph $\emptyset\varsubsetneq\gamma\varsubsetneq\Gamma$ is said \textbf{shrinkable} if $\res(\gamma)\in R_{V}$. The set of shrinkable subgraphs of $\Gamma$ will be denoted by $\mathbf{\Gsub}$.
\end{defn}
Note that until now we did not really define what is the map $\res$. We now do it. First we assume that it is an algebra homomorphism from $\cH$ to $\cH\cup R_{V}$. Then to compute the graphical residue of a generator of $\cH$, we need the following remarks and definitions.

The coproduct of $\cH$ (usually given by (\ref{eq:coproduct1})) drives the combinatorial and algebraic aspects of renormalization if it corresponds to some analytical facts. Before we explain this, let us recall the following definitions.
\begin{defn}\label{def:FeynmanRules}
  The (unrenormalized) \textbf{Feynman rules} are an homomorphism $\mathbf{\phi}$ from $\cH$ to $\cA$. The precise definition of $\cA$ depends on the regularization scheme employed (in dimensional regularization, $\cA$ is the Laurent series).
\end{defn}
\begin{defn}\label{def:projectionT}
  The \textbf{projection} $\mathbf{T}$ is a map from $\cA$ to $\cA$ wich has to fulfill: $\forall\Gamma\in\cH$, $\Gamma$ primitive
  \begin{align}
    (\id_{\cA}-T)\circ\phi(\Gamma)<\infty.\label{eq:conditionT}
  \end{align}
  This means that if $\phi(\Gamma)$ is superficially divergent (as the cut-off is removed) then its overall divergence is totally included in $T\circ\phi(\Gamma)$.%  This also means that up to some finite contributions, $\phi(\Gamma)$ behaves like $T\circ\phi(\Gamma)$.
\end{defn}

\paragraph{External structures}
The projection $T$ extracts the divergent part of the amplitude $\phi(\Gamma)$. In the case of a two-point graph this divergent part decomposes into two pieces. The first one is a mass term whereas the second one contributes to the wave function renormalization (recall that the propagator of the commutative $\phi^{4}$ theory is $(-\Delta+m^{2})^{-1}$). To distinguish between these two, one introduces \emph{external structures} \cite{Connes:2000uq,Kreimer:2005lr}. It consists in the following endomorphisms of $\cA$ (in $x$-space representation):
\begin{subequations}
  \label{eq:ExtStruct}
  \begin{align}
    \langle\sigma_{0},\phi(\Gamma)\rangle=&\rho_{0}(\Gamma)\delta_{y}(x),\label{eq:ExtStructMass}\\
    \langle\sigma_{1},\phi(\Gamma)\rangle=&\rho_{1}(\Gamma)\Delta\delta_{y}(x),\label{eq:ExtStructWave}\\
    \langle\sigma_{2},\phi(\Gamma)\rangle=&\rho_{2}(\Gamma)\delta_{x_{2}}(x_{1})\delta_{x_{3}}(x_{1})\delta_{x_{4}}(x_{1})\label{eq:ExtStructVertex}
  \end{align}
\end{subequations}
where the $\rho_{i}$'s are characters on $\cA$. If $K_{\Gamma}$ is the kernel of the amplitude $\phi(\Gamma)$, these characters are given by:
\begin{subequations}
  \label{eq:Rho}
  \begin{align}
    \rho_{0}(\Gamma)=&\int d^{4}z\,K_{\Gamma}(x,z),\label{eq:Rho0}\\
    \rho_{1}(\Gamma)=&\frac 18\int d^{4}z\,(z-x)^{2}K_{\Gamma}(x,z),\label{eq:Rho1}\\
    \rho_{2}(\Gamma)=&\int d^{4}x_{2}d^{4}x_{3}d^{4}x_{4}\,K_{\Gamma}(x,x_{2},x_{3},x_{4}).\label{eq:Rho2}
  \end{align}
\end{subequations}
Recall that commutative field theories are usually translation invariant so that none of the $\rho_{i}$'s depend on $x$. With those notations, $T=\sigma_{0}+\sigma_{1}$ on a two-point graph and $T=\sigma_{2}$ on a four-point graph.\\

There is now a way to relate the analytical operations $\sigma_{i}$'s to the graphical map $\res$:
\begin{defn}[Residue]\label{def:residue}
  The residue map $\res : \cH\to\cH\cup R_{V}$ is defined by
  \begin{align}
    \langle\sigma_{i},\phi(\Gamma)\rangle=&\rho_{i}(\Gamma)\langle\sigma_{i},\phi\circ\res(\Gamma)\rangle\label{eq:resdef}
  \end{align}
  where $i=0\text{ or }1$ for a two-point graph and $i=2$ on a four-point graph.
\end{defn}
Following equations (\ref{eq:ExtStruct}) and (\ref{eq:Rho}) one finds
\begin{subequations}
  \label{eq:AnaRes}
  \begin{align}
    \phi\circ\res(\Gamma)=&\delta_{y}(x)+\Delta\delta_{y}(x)&&\text{if $\Gamma$ is a two-point graph,}\label{eq:AnaRes2}\\
    \phi\circ\res(\Gamma)=&\delta_{x_{2}}(x_{1})\delta_{x_{3}}(x_{1})\delta_{x_{4}}(x_{1})&&\text{if $\Gamma$ is a four-point graph}\label{eq:AnaRes4}
  \end{align}
  which leads to the following graphical definitions:
  \begin{align}
    \res(~\parbox{20pt}{
      \begin{fmfgraph}(20,10)
        \fmfleft{l}
        \fmfright{r}
        \fmf{plain}{l,v}
        \fmf{plain}{r,v}
        \fmfv{decor.shape=circle, decor.filled=shaded, decor.size=3.5thick}{v}
      \end{fmfgraph}}~)=\parbox{20pt}{
      \begin{fmfgraph}(20,10)
        \fmfleft{l}
        \fmfright{r}
        \fmf{plain}{l,r}
      \end{fmfgraph}}=\big(\parbox{20pt}{
      \begin{fmfgraph*}(20,10)
        \fmfleft{l}
        \fmfright{r}
        \fmf{plain}{l,v}
        \fmf{plain}{r,v}
        % \fmfdot{v}
        % \fmfblob{1.5thick}{v}
        \fmfv{decor.shape=circle, decor.filled=full, decor.size=1.0thick}{v}
        \fmfv{label={\tiny{$0$}},label.angle=90,label.dist=1.5mm}{v}
      \end{fmfgraph*}}~+~
    \parbox{20pt}{
      \begin{fmfgraph*}(20,10)
        \fmfleft{l}
        \fmfright{r}
        \fmf{plain}{l,v}
        \fmf{plain}{r,v}
        % \fmfdot{v}
        % \fmfblob{1.5thick}{v}
        \fmfv{decor.shape=circle, decor.filled=full, decor.size=1.0thick}{v}
        % \fmfv{decor.shape=cross, decor.size=2.5thick}{v}
        \fmfv{label={\tiny{$1$}},label.angle=90,label.dist=1.5mm}{v}
      \end{fmfgraph*}}\big)^{-1}~,~~
    \res(~\parbox{25pt}{
      \begin{fmfgraph}(20,10)
        \fmfleft{e1,e2}
        \fmfright{e3,e4}
        \begin{fmffor}{n}{1}{1}{4}
          \fmf{plain}{e[n],v}
        \end{fmffor}
        \fmfv{decor.shape=circle, decor.filled=shaded, decor.size=3.5thick}{v}
      \end{fmfgraph}})=\parbox{20pt}{
      \begin{fmfgraph}(20,10)
        \fmfleft{l1,l2}
        \fmfright{r1,r2}
        \fmf{plain}{l1,v,l2}
        \fmf{plain}{r1,v,r2}
        \fmfv{decor.shape=circle, decor.filled=full, decor.size=1thick}{v}
      \end{fmfgraph}}~.
  \end{align}
\end{subequations}

% Let us make a couple of remarks concerning the last two definitions. The condition (\ref{eq:conditionT}) means that the map $T$ really captures the divergent part of the amplitude of a graph. Note that $T\circ\phi(\Gamma)$ may also contain convergent contributions depending on the substraction scheme precisely fixed by the choice of $T$.
Equation (\ref{eq:resdef}) means that the divergent part of a graph $\Gamma$ ``looks like'' another graph called $\res(\Gamma)$. For a renormalizable quantum field theory the residue of any superficially divergent graph belongs to $R_{V}$. This is the usual statement according to which all the divergences of a renormalizable field theory can be ``absorbed'' in a redefinition of the various coupling constants. If the theory is local then $\res(\Gamma)$ corresponds to the graph obtained from $\Gamma$ by shrinking all its internal lines to a point. But this is a particular case and we have to define $\res$ as reflecting the appropriate projection $T$. For example, we will see in the next section that the residue of a non-commutative graph is not a local graph anymore.

The $T$ operation is designed to extract the ``main'' part of graphs. For the convergent ones there is no good distinction between $T\circ\phi(\Gamma)$ and $(\id-T)\circ\phi(\Gamma)$: both are convergent expressions. That's why $T$ is mainly defined on (superficially) divergent graphs. Nevertheless one can define $T$ to be $\id_{\cA}$ on convergent graphs. Condition (\ref{eq:conditionT}) is then trivially fulfilled and equation (\ref{eq:resdef}) is satisfied with $\res=\id_{\cH}$ and $\rho$ the trivial character.

\subsection{Coassociative coproducts}
\label{sec:coasociative-coproducts}

Using the definitions of section \ref{sec:residue-map} we have the following lemma:
\begin{lemma}[Coassociativity]\label{lem:coassociative1}
  Let $\Gamma\in\cH$. Provided
  \begin{enumerate}
  \item $\forall\gamma\in\Gsub,\,\forall\gamma'\in\sub{\gamma}$ such that $\res(\gamma)\in R_{V}$ and $\res(\gamma')\in R_{V}$, $\res(\gamma/\gamma')\in R_{V}$,\label{item:condition1}
  \item $\forall\gamma_{1}\in\cH,\,\forall\gamma_{2}\in\cH$ such that $\res(\gamma_{1})\in R_{V}$ and $\res(\gamma_{2})\in R_{V}$, there exists gluing data $G$ such that $\res(\gamma_{1}\circ_{G}\gamma_{2})\in R_{V}$,\label{item:condition2}
  \end{enumerate}
  the following coproduct is coassociative
  \begin{subequations}
    \label{eq:coproduct1}
    \begin{align}
      \Delta\Gamma=&\Gamma\otimes\bbbone+\bbbone\otimes\Gamma+\Delta'\Gamma,\label{eq:Deltaprime}\\
      \Delta'\Gamma=&\sum_{\gamma\in\Gsub}\gamma\otimes\Gamma/\gamma.
    \end{align}
  \end{subequations}
\end{lemma}
Remark that $\Gamma/\gamma$ is the graph obtained from $\Gamma$ by replacing $\gamma\subset\Gamma$ by its residue. Then $\res(\gamma)\in R_{V}$ implies $\Gamma/\gamma\in\cH$. We prove this lemma by following closely \cite{Connes:2000uq}.
\begin{proof}
  First note that $(\Delta\otimes\id)\Delta=(\id\otimes\Delta)\Delta\iff (\Delta'\otimes\id)\Delta'=(\id\otimes\Delta')\Delta'$ which means that all the following subgraphs can be considered as neither full nor empty. Let $\Gamma$ a generator of $\cH$,
  \begin{align}
    (\Delta'\otimes\id)\Delta'\Gamma=&(\Delta'\otimes\id)\sum_{\gamma\in\Gsub}\gamma\otimes\Gamma/\gamma\\
    =&\sum_{\gamma\in\Gsub}\sum_{\gamma'\in\sub{\gamma}}\gamma'\otimes\gamma/\gamma'\otimes\Gamma/\gamma\label{eq:coprodproofLeft1}\\
    (\id\otimes\Delta')\Delta'\Gamma=&\sum_{\gamma'\in\Gsub}\sum_{\gamma''\in\sub{\Gamma/\gamma'}}\gamma'\otimes\gamma''\otimes(\Gamma/\gamma')/\gamma''.\label{eq:coprodproofRight1}
  \end{align}
  By the definitions \ref{def:subgraph}, \ref{def:shrinksubgraph} and \ref{def:residue} it is clear that $\gamma'\in\sub{\gamma}$ and $\gamma\in\Gsub$ implies $\gamma'\in\Gsub$. This implicitly uses the fact that the residue of a graph is independant of the surrounding of this graph and really only depends on the graph itself: $\res(\gamma)$ is the same wether $\gamma$ is a subgraph or not of another graph. Equation (\ref{eq:coprodproofLeft1}) can then be rewritten as
  \begin{align}
    (\Delta'\otimes\id)\Delta'\Gamma=&\sum_{\gamma'\in\Gsub}\;\sum_{\gamma\in\Gsub\tq\gamma\varsupsetneq\gamma'}\gamma'\otimes\gamma/\gamma'\otimes\Gamma/\gamma.\label{eq:coprodproofLeft2}
  \end{align}
  It is now enough to prove equality between (\ref{eq:coprodproofRight1}) and (\ref{eq:coprodproofLeft2}) at fixed $\gamma'\in\Gsub$. Let us first fix a subgraph $\gamma\in\Gsub$ such that $\gamma\varsupsetneq\gamma'$ and prove that there exists a graph $\gamma''\in\sub{\Gamma/\gamma'}$ such that $\gamma/\gamma'\otimes\Gamma/\gamma=\gamma''\otimes(\Gamma/\gamma')/\gamma''$. Of course the logical choice for $\gamma''$ is $\gamma/\gamma'$ because then $(\Gamma/\gamma')/(\gamma/\gamma')=\Gamma/\gamma$.

We only have to prove that $\gamma''=\gamma/\gamma'\in\sub{\Gamma/\gamma'}$. It is clear that $\gamma/\gamma'$ is a subset of internal lines of $\Gamma/\gamma'$. Then $\gamma/\gamma'\in\sub{\Gamma/\gamma'}$ if $\res(\gamma)\in R_{V}$ and $\res(\gamma')\in R_{V}$ implies $\res(\gamma/\gamma')\in R_{V}$ which we assumed.

Conversely let us fix $\gamma''\in\sub{\Gamma/\gamma'}$ and prove that there exists $\gamma\in\Gsub$ containing $\gamma'$ such that $\gamma/\gamma'\otimes\Gamma/\gamma=\gamma''\otimes(\Gamma/\gamma')/\gamma''$. Let us write $\gamma'=\bigcup_{i\in I}\gamma'_{i}$ for the connected components of $\gamma'$. Some of these components led to vertices of $\gamma''$, the others to vertices of $(\Gamma/\gamma')\setminus\gamma''$. We can then define $\gamma$ as $(\gamma''\circ_{G_{I_{1}}}\bigcup_{i\in I_{1}}\gamma'_{i})\bigcup_{j\in I_{2}}\gamma'_{j}$ with $I_{1}\cup I_{2}=I$. It is clearly a subgraph of $\Gamma$ and belongs to $\Gsub$ if $\forall \gamma_{1},\gamma_{2}\in\cH,\,\res(\gamma_{1})\in R_{V},\,\res(\gamma_{2})\in R_{V}\text{ there exists gluing data } G \text{ such that }\res(\gamma_{1}\circ_{G}\gamma_{2})\in R_{V}$. We also assumed it. This ends the proof of Lemma \ref{lem:coassociative1}.
\end{proof}

Let us now work out how Lemma \ref{lem:coassociative1} fits the commutative $\phi^{4}$ model. In this local field theory the divergent graphs have two or four external legs. The residue of a given graph is the one obtained by shrinking all its internal lines to a point (see section \ref{sec:residue-map}) and then only depends on the number of external lines of the graph. Let us check condition \ref{item:condition2} of Lemma \ref{lem:coassociative1} for commutative $\phi^{4}$. We consider two graphs $\gamma_{1}$ and $\gamma_{2}$ with two or four external legs. We consider $\gamma_{0}=\gamma_{1}\circ_{G}\gamma_{2}$ for any gluing data $G$. Let $V_{i},\,I_{i}$ and $E_{i}$ the respective numbers of vertices, internal and external lines of $\gamma_{i},\,i\in\{0,1,2\}$. For all $i\in\{0,1,2\}$, we have
\begin{subequations}
  \label{eq:ConservationExtLines}
  \begin{align}
    4V_{i}=&2I_{i}+E_{i}\\
    V_{0}=&
    \begin{cases}
      V_{1}+V_{2}&\text{if $E_{2}=2$}\\
      V_{1}+V_{2}-1&\text{if $E_{2}=4$}
    \end{cases}\\
    I_{0}=&
    \begin{cases}
      I_{1}+I_{2}+1&\text{if $E_{2}=2$}\\
      I_{1}+I_{2}&\text{if $E_{2}=4$}
    \end{cases}
  \end{align}
\end{subequations}
which proves that $E=E_{1}$. Then as soon as $\res(\gamma_{1})\in R_{V}$ so does $\res(\gamma_{0})$. Concerning condition \ref{item:condition1} note that $\gamma''=\gamma/\gamma'\Longleftrightarrow\exists G\tq\gamma=\gamma''\circ_{G}\gamma'$ which allows to prove, in the case of a local theory, that condition \ref{item:condition1} also holds and that the coproduct (\ref{eq:coproduct1}) is coassociative.

\begin{lemma}
  Let $\mathbf{\cH_{c}}$ the linear space of graphs whose residue is $R_{V}$-valued:
  \begin{align}
    \cH_{c}=&\lb\Gamma\in\cH\tqs\res(\Gamma)\in R_{V}\rb.
  \end{align}
  $\cH_{c}$ is a Hopf subalgebra of $\cH$.
\end{lemma}
\begin{proof}
  Thanks to the definition (\ref{eq:coproduct1}), $\Delta\cH_{c}\subset\cH_{c}\otimes\cH_{c}$. By induction on the augmentation degree, one also proves that $S(\cH_{c})\subset\cH_{c}$.
%   \begin{align}
%     S(\Gamma)=&-\Gamma-\sum_{\gamma\in\Gsub}S(\gamma)\Gamma/\gamma.    
%   \end{align}
\end{proof}

% \begin{com}
%   For NCFT, $\cH_{c}$ is not the algebra of divergent graphs: cf. orientable $B=1$ non-planar graphs.
% \end{com}

\section{Hopf algebra for non-commutative Feynman graphs}
\label{sec:hopf-algebra-nonc}

The definition of the Hopf algebra of non-commutative Feynamn graphs which drives the combinatorics of perturbative renormalization is formally the same as in the commutative case \cite{Connes:2000uq}. But before giving the definitions let us define the residue of a \encv{} graph. As already mentionned it has been proven (first in \cite{GrWu04-3}) that the Grosse-Wulkenhaar model \eqref{action} is renormalizable to all orders of perturbation. It means that the divergent parts of the divergent graphs are proportionnal to mass, wave-function, $x^{2}$ and Moyal vertex terms. Following the procedure exposed in section \ref{sec:residue-map}, particularly equations \eqref{eq:resdef} and \eqref{eq:AnaRes}, we find
\begin{subequations}
  \label{eq:NCAnaRes}
  \begin{align}
    \phi\circ\res(\Gamma)=&\delta_{y}(x)+\Delta\delta_{y}(x)+\xt^{2}\delta_{y}(x)&&\text{if $\Gamma$ is a two-point planar regular graph,}\label{eq:NCAnaRes2}\\
    \phi\circ\res(\Gamma)=&\big(\delta_{x_{2}}\star\delta_{x_{3}}\star\delta_{x_{4}}\big)(x_{1})&&\text{if $\Gamma$ is a four-point planar regular graph}\label{eq:NCAnaRes4}
  \end{align}
  which leads to the following graphical definitions:
  \begin{align}
    \res(~\parbox{20pt}{
      \begin{fmfgraph}(20,10)
        \fmfleft{l}
        \fmfright{r}
        \fmf{plain}{l,v}
        \fmf{plain}{r,v}
        \fmfv{decor.shape=circle, decor.filled=shaded, decor.size=3.5thick}{v}
      \end{fmfgraph}}~)=\parbox{20pt}{
      \begin{fmfgraph}(20,10)
        \fmfleft{l}
        \fmfright{r}
        \fmf{plain}{l,r}
      \end{fmfgraph}}~,~~
    \res(~\parbox{25pt}{
      \begin{fmfgraph}(20,10)
        \fmfleft{e1,e2}
        \fmfright{e3,e4}
        \begin{fmffor}{n}{1}{1}{4}
          \fmf{plain}{e[n],v}
        \end{fmffor}
        \fmfv{decor.shape=circle, decor.filled=shaded, decor.size=3.5thick}{v}
      \end{fmfgraph}})=\parbox{20pt}{\includegraphics[scale=.3]{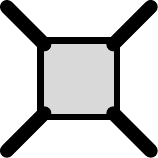}}~.
  \end{align}
\end{subequations}
Once more the (graphical) residue of a convergent graph is defined as $\id_{\cH}$.

Consider now the unital associative algebra  $\cH$ freely generated by $1$PI non-commutative Feynman graphs (including the empty set, which we denote by $\bbbone$). The product $m$ is bilinear, commutative and given by the operation of disjoint union. Let the coproduct $\Delta:\cH\to\cH\otimes\cH$ defined as
\begin{equation}
\Delta \Gamma = \Gamma \otimes\bbbone + \bbbone\otimes \Gamma + \sum_{\gamma\in \Gsub} \gamma \otimes \Gamma/\gamma, \ \forall\Gamma\in\cH.\label{eq:NCcoproduct}
\end{equation}
Furthermore let us define the counit $\veps:\cH\to\K$:
\begin{equation}
\veps (\bbbone) =1,\ \veps (\Gamma)=0,\ \forall \Gamma\ne\bbbone.
\end{equation}
Finally the antipode is given recursively by
\begin{align}
  S:\cH\to&\cH\label{eq:Antipode}\\
  \Gamma\mapsto&-\Gamma-\sum_{\gamma\in\Gsub}S(\gamma)\Gamma/\gamma.\nonumber
\end{align}
We can state the main result of this letter:
\begin{thm}\label{thm:hopf-algebra-nonComm}
The quadruple $(\cH,\Delta,\veps,S)$ is a Hopf algebra.
\end{thm}
\begin{proof}
  The only thing to prove is the coassociativity of the coproduct \eqref{eq:NCcoproduct}. Once it is done, the definition \eqref{eq:Antipode} for the antipode follows from the fact that $\cH$ is graded (by the loop number), connected and from Lemma \ref{lem:Sfree}.

We will use Lemma \ref{lem:coassociative1} and the fact that for all $\Gamma\in\cH$, $\res(\Gamma)\in R_{V}$ is equivalent to $\Gamma$ is planar regular. Then conditions \ref{item:condition1} and \ref{item:condition2} of Lemma \ref{lem:coassociative1} are equivalent to:
\begin{enumerate}
\item for all $\gamma$ and $\gamma'\subset\gamma$ both planar regular, $\gamma/\gamma'$ is planar regular,\label{item:cond1proof}
\item for all $\gamma$ and $\gamma'\subset\gamma$ both planar regular, there exits gluing data $G$ such that $\gamma\circ_{G}\gamma'$ is planar regular.\label{item:cond2proof}
\end{enumerate}
In the following all the graphs we are going to insert will be four-point graphs. The case of two-point graphs is easier and left to the reader. Before proving conditions \ref{item:cond1proof} and \ref{item:cond2proof}, let us consider the insertion of a regular four-point graph $\gamma_{2}$ into a vertex of another graph $\gamma_{1}$.
\begin{figure}[htbp]
  \centering 
  \subfloat[A vertex of $\gamma_{1}$]{{\label{fig:vert1}}\raisebox{1.3cm}{\makebox[4cm]{\includegraphics[scale=1]{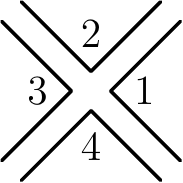}}}}\hspace{1cm}\qquad
  \subfloat[Insertion of $\gamma_{2}$]{\label{fig:insertion}\includegraphics[scale=1]{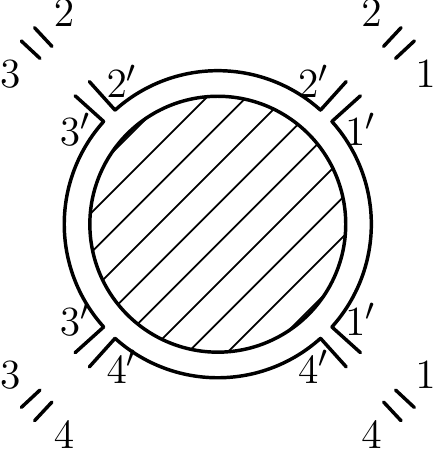}}
  \caption{Insertion procedure}
  \label{fig:InsertProc}
\end{figure}
Let $\gamma_{0}=\gamma_{1}\circ\gamma_{2}$ and for all $i\in\{0,1,2\}$ let $F_{i},I_{i},V_{i},B_{i}$ the respective numbers of faces, internal lines, vertices and broken faces of $\gamma_{i}$. The number of faces of a ribbon graph is the number of closed\footnote{In the case of external faces, one considers that the corresponding lines are closed.} single lines. A ribbon vertex is drawn on figure \ref{fig:vert1}. One sees that the number of faces to which the lines of that vertex belong is at most four. Some of them may indeed belong to the same face. The gluing data necessary to the insertion of $\gamma_{2}$ corresponds to a bijection between the half-lines of the vertex in $\gamma_{1}$ and the external lines of $\gamma_{2}$. This last one being regular (only one broken face) the typical situation is represented on figure \ref{fig:insertion}. It should be clear that $F=F_{2}-1+F_{1}-n$ for some $n\ges 0$. $F_{2}-1$ is the number of internal faces of $\gamma_{2}$ i.e. the number of faces of the blob. The number $n$ depends on the gluing data. It vanishes if the insertion respects the \emph{cyclic ordering} of the vertex. For example the following bijection $\sigma$ does:
\begin{align}
  \sigma((1',2'))=&(2,3),\ \sigma((2',3'))=(3,4),\ \sigma((3',4'))=(4,1),\ \sigma((4',1'))=(1,2).  
\end{align}
As in equations \eqref{eq:ConservationExtLines}, $I_{0}=I_{1}+I_{2}$ and $V_{0}=V_{1}+V_{2}-1$. It follows that the genus of $\gamma_{0}$ satisfies
\begin{align}
  g(\gamma_{0})=&g(\gamma_{1})+g(\gamma_{2})+n.\label{eq:GenusInsert}
\end{align}
Moreover by exhausting the $4!/4$ possible insertions, one checks that $B_{0}\ges B_{1}$. For example, on figure \ref{fig:ExtSit}, lines $1$ and $3$ belong to two different broken faces. Figure \ref{fig:InsRegGraph} shows an insertion of a regular four-point graph which increases the number of broken faces by one: now trajectories $(1,4),(1,2)$ and $3$ are external faces (line $(2,4)$ is still an internal one).
\begin{figure}[htbp]
  \centering 
  \subfloat[External situation]{\label{fig:ExtSit}\includegraphics[scale=1]{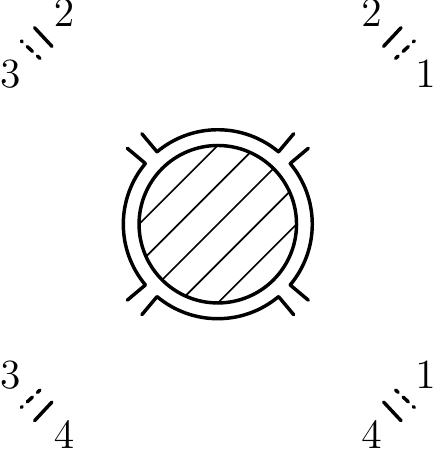}}\hspace{1cm}\qquad
  \subfloat[Insertion of a regular graph]{\label{fig:InsRegGraph}\makebox[6cm]{\includegraphics[scale=1]{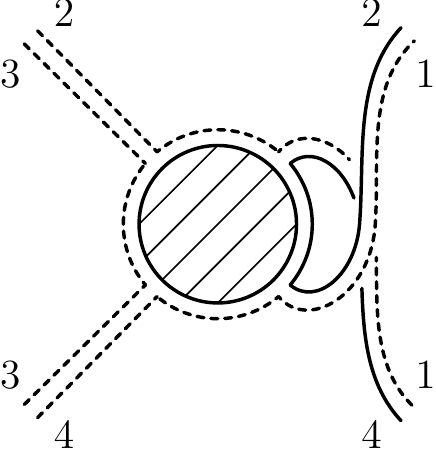}}}
  \caption{Increasing number of external faces}
  \label{fig:InsertProcB}
\end{figure}

Let us now turn to proving that the algebra of non-commutative Feynman graphs described above fulfills conditions \ref{item:cond1proof} and \ref{item:cond2proof}.
\begin{enumerate}
\item $\gamma,\gamma'$ planar implies $\gamma/\gamma'$ planar thanks to equation \eqref{eq:GenusInsert}. Furthermore $B(\gamma)=1$ implies $B(\gamma/\gamma')=1$ due to the preceding remark.
\item For condition \ref{item:cond2proof} one chooses gluing data $G$ respecting the cyclic ordering of the vertex. Then one has $g(\gamma\circ_{G}\gamma')=g(\gamma)+g(\gamma')=0$. The cyclic ordering of the insertion ensures $B(\gamma\circ_{G}\gamma')=B(\gamma)=1$.
\end{enumerate}
\end{proof}

Let $f,g\in\text{Hom}(\cH,\cA)$ where $\cA$ is the range algebra of the projection $T$ (see subsection \ref{sec:residue-map}). The convolution product $\ast$ in $\text{Hom}(\cH,\cA)$ is defined by
\begin{align}
  f\ast g=&m_{\cA}\circ(f\otimes g)\circ\Delta_{\cH}.\label{eq:convolution}
\end{align}
Let $\phi$ the unrenormalized Feynman rules and $\phi_{-}\in\text{Hom}(\cH,\cA)$ the twisted antipode: $\forall\Gamma\in\cH$,
\begin{align}
  \phi_{-}(\Gamma)=&-T\big(\phi(\Gamma)+\sum_{\gamma\in\Gsub}\phi_{-}(\gamma)\ \phi(\Gamma/\gamma)\big).\label{eq:PhiMinus}
  \end{align}
As in the commutative field theories, the renormalized amplitude $\phi_{+}$ of a graph $\Gamma\in\cH$ is given by:
\begin{align}
  \phi_{+}(\Gamma)=&\phi_{-}\ast\phi(\Gamma).\label{eq:RenormValue}
  \end{align}

\end{fmffile}

\paragraph{Acknowledgment}We would like to warmly thank Dirk Kreimer for fruitful discussions.

\bibliographystyle{fabalpha-en}
\bibliography{biblio-articles,biblio-books}

\end{document}